
\input phyzzx
%
%
\catcode`@=11 
%
%
%
%
\newfam\ssfam   
\font\seventeenss =cmss10 scaled\magstep4
\font\fourteenss  =cmss10 scaled\magstep2
\font\twelvess    =cmss10 scaled\magstep1
\font\tenss       =cmss10
\font\niness      =cmss9
\font\eightss     =cmss8
\def\seventeenpoint{\relax
    \textfont0=\seventeenrm          \scriptfont0=\twelverm
      \scriptscriptfont0=\ninerm
    \textfont1=\seventeeni           \scriptfont1=\twelvei
      \scriptscriptfont1=\ninei
    \textfont2=\seventeensy          \scriptfont2=\twelvesy
      \scriptscriptfont2=\ninesy
    \textfont3=\seventeenex          \scriptfont3=\twelveex
      \scriptscriptfont3=\ninex
    \textfont\itfam=\seventeenit    
    \textfont\slfam=\seventeensl    
      \scriptscriptfont\slfam=\ninesl
    \textfont\bffam=\seventeenbf     \scriptfont\bffam=\twelvebf
      \scriptscriptfont\bffam=\ninebf
    \textfont\ttfam=\seventeentt
    \textfont\cpfam=\seventeencp
    \textfont\ssfam=\seventeenss     \scriptfont\ssfam=\twelvess
      \scriptscriptfont\ssfam=\niness
    \samef@nt
    \b@gheight=17pt
    \setbox\strutbox=\hbox{\vrule height 0.85\b@gheight
                                depth 0.35\b@gheight width\z@ }}
\def\fourteenf@nts{\relax
    \textfont0=\fourteenrm          \scriptfont0=\tenrm
      \scriptscriptfont0=\sevenrm
    \textfont1=\fourteeni           \scriptfont1=\teni
      \scriptscriptfont1=\seveni
    \textfont2=\fourteensy          \scriptfont2=\tensy
      \scriptscriptfont2=\sevensy
    \textfont3=\fourteenex          \scriptfont3=\twelveex
      \scriptscriptfont3=\tenex
    \textfont\itfam=\fourteenit     \scriptfont\itfam=\tenit
    \textfont\slfam=\fourteensl     \scriptfont\slfam=\tensl
    \textfont\bffam=\fourteenbf     \scriptfont\bffam=\tenbf
      \scriptscriptfont\bffam=\sevenbf
    \textfont\ttfam=\fourteentt
    \textfont\cpfam=\fourteencp
    \textfont\ssfam=\fourteenss     \scriptfont\ssfam=\tenss
        \scriptscriptfont\ssfam=\sevenrm }
\def\twelvef@nts{\relax
    \textfont0=\twelverm          \scriptfont0=\ninerm
      \scriptscriptfont0=\sixrm
    \textfont1=\twelvei           \scriptfont1=\ninei
      \scriptscriptfont1=\sixi
    \textfont2=\twelvesy           \scriptfont2=\ninesy
      \scriptscriptfont2=\sixsy
    \textfont3=\twelveex          \scriptfont3=\tenex
      \scriptscriptfont3=\tenex
    \textfont\itfam=\twelveit     \scriptfont\itfam=\nineit
    \textfont\slfam=\twelvesl     \scriptfont\slfam=\ninesl
    \textfont\bffam=\twelvebf     \scriptfont\bffam=\ninebf
      \scriptscriptfont\bffam=\sixbf
    \textfont\ttfam=\twelvett
    \textfont\cpfam=\twelvecp    \scriptfont\cpfam=\tencp
    \textfont\ssfam=\twelvess    \scriptfont\ssfam=\niness
      \scriptscriptfont\ssfam=\sixrm }
\def\tenf@nts{\relax
    \textfont0=\tenrm          \scriptfont0=\sevenrm
      \scriptscriptfont0=\fiverm
    \textfont1=\teni           \scriptfont1=\seveni
      \scriptscriptfont1=\fivei
    \textfont2=\tensy          \scriptfont2=\sevensy
      \scriptscriptfont2=\fivesy
    \textfont3=\tenex          \scriptfont3=\tenex
      \scriptscriptfont3=\tenex
    \textfont\itfam=\tenit     \scriptfont\itfam=\seveni  
    \textfont\slfam=\tensl     \scriptfont\slfam=\sevenrm 
    \textfont\bffam=\tenbf     \scriptfont\bffam=\sevenbf
      \scriptscriptfont\bffam=\fivebf
    \textfont\ttfam=\tentt
    \textfont\cpfam=\tencp
    \textfont\ssfam=\tenss      \scriptfont\ssfam=\eightss
      \scriptscriptfont\ssfam=\fiverm }
\def\ss{\n@expand\f@m\ssfam}
\font\bfss=cmssbx10
\font\fortssbx=cmssbx10 scaled \magstep2
%
\newdimen\madphheadsize \madphheadsize=7.0in
\newdimen\madphheadleft \madphheadleft=3.75in
\newdimen\madpheadsize \madpheadsize=7.0in
\newdimen\madpheadleft \madpheadleft=3.5in
\showboxbreadth=1000 %
\showboxdepth=5
\newdimen\madphheadsize \madphheadsize=7.0in
\newdimen\madphheadleft \madphheadleft=3.6875in
\def\MADPHHEAD{{\Tenpoint\vbox{\vskip-0.7in
  \line{\hss
   \hbox{\vbox{\hbox to \madphheadsize{\hskip\madphheadleft
                        \fortssbx  University of Wisconsin - Madison\hfil}
               \vskip 5pt
               \hbox to \madphheadsize{\hskip\madphheadleft
                         \bfss Department of Physics\hfil}
               \vskip -45pt
	       \hbox to \madphheadsize{\hskip 3.25in
			{\logo A}
                          \hfill}
               \vskip 24pt \hrule height 1pt \vskip 6pt
               \hbox to \madphheadsize{\hbox to 0pt{\hskip 1pt
                  \tenss High Energy Physics\hss}
    \hskip \madphheadleft
    \tenss 1150 University Avenue, Madison, Wisconsin 53706\hfil}
   \hbox to \madphheadsize{\hbox to 0pt{\hskip 1pt
    \tenss Telephone: 608/262-2281\hss}
    \hskip \madphheadleft
    \tenss Telex: 265452 UOFWISC MDS\hfil}}}\hss}}}}

\begingroup
 \catcode `\{ = 12  
 \catcode `\} = 12  
 \catcode `\[ = 1
 \catcode `\] = 2
 \gdef\labelformlabels[%
   \gdef\rwl@begin##1\cr[\rw@toks=[##1]\rel@x
        \immediate\write\labelswrite[\the\rw@toks]\futurelet\n@xt\rwl@next]
   \gdef\writenextlabel##1[%
        \immediate\write\labelswrite[  ]%
        \immediate\write\labelswrite[{]%
            \rwl@begin ##1%
            \rwl@end%
        \immediate\write\labelswrite[}]]%
   \gdef\writelabel##1[%
        \immediate\write\labelswrite[{]%
            \rwl@begin ##1%
            \rwl@end%
        \immediate\write\labelswrite[}]%
        \let\writelabel=\writenextlabel]%
]
\endgroup
%
%

%
%
%
\def\figitem#1{\r@fitem{#1.}}
\def\tabitem#1{\r@fitem{#1.}}

\def\sequentialequations{\rel@x \ifnum\equanumber<0 \else
  \gl@bal\equanumber=-\equanumber \gl@bal\advance\equanumber by -1 \fi }

\def\boxit#1{\vbox{\hrule\hbox{\vrule\kern3pt
\vbox{\kern3pt#1\kern3pt}\kern3pt\vrule}\hrule}}
%
%
\newbox\partialpage
\newdimen\pageheight \pageheight=\vsize
\newdimen\pagewidth  \pagewidth=6.6truein
\newdimen\columnwidth  \columnwidth=3.2truein
\newdimen\columnrulewidth \columnrulewidth=0pt
\newdimen\ruleht \ruleht=.5pt
\newinsert\margin
\def\twocolumn{%
   \singlespace
   \vsize=9truein
   \pagetextwidth=\pagewidth
   \hsize=\pagewidth
   \titlepagewidth=\pagewidth
   \hoffset=0truein
   \voffset=0truein
   \dimen\margin=\maxdimen
   \count\margin=0 \skip\margin=0pt
 \def\begindoublecolumns{
     \ifpr@printstyle
     \message{ I'm unable to print double columns in PREPRINTSTYLE }
     \end\fi
     \begingroup
     \global\vsize=2\pageheight
     \output={\global\setbox\partialpage=\vbox{\unvbox255\bigskip\bigskip}
         \global\vsize=2\pageheight\global\advance\vsize by -2\ht\partialpage
         \global\advance\vsize by 2\bigskipamount
         \global\advance\vsize by 1 pc}\eject 
     \output={\doublecolumnout\global\vsize=2\pageheight}
         \global\pagetextwidth=\columnwidth \global\hsize=\columnwidth}
%
  \def\enddoublecolumns{\output={\balancecolumns\global\hsize=\pagewidth
                       \global\pagetextwidth=\pagewidth
                       \global\vsize=\pageheight
                       \unvbox255 }\eject\endgroup}
  \def\doublecolumnout{\splittopskip=\topskip \splitmaxdepth=\maxdepth
     \dimen@=\pageheight\advance\dimen@ by -\ht\partialpage
     \setbox0=\vsplit255 to\dimen@ \setbox2=\vsplit255 to \dimen@
     \onepageout\pagesofar \unvbox255 \penalty\outputpenalty}
  \def\pagesofar{\unvbox\partialpage
     \wd0=\columnwidth \wd2=\columnwidth \hbox to \pagewidth{\box0\hfil
     \columnrule \hfil \box2}}
  \def\columnrule{\vrule width \columnrulewidth height \ht2}
  \def\balancecolumns{\setbox0=\vbox{\unvbox255}\dimen@=\ht0
     \advance\dimen@ by \topskip \advance\dimen@ by-\baselineskip
     \advance\dimen@ by -2\ht\partialpage  
     \divide\dimen@ by2                    
     \splittopskip=\topskip
     {\vbadness=10000 \loop \global\setbox3=\copy0
        \global\setbox1=\vsplit3 to \dimen@
        \ifdim\ht3>\dimen@ \global\advance\dimen@ by1pt \repeat}
     \setbox0=\vbox to \dimen@{\unvbox1} \setbox2=\vbox to \dimen@{\dimen2=\dp3
     \unvbox3 \kern-\dimen2 \vfil }
     \pagesofar }
   \def\onepageout##1{ \setbox0=\vbox{##1} \dimen@=\dp0
     \shipout\vbox{ 
     \makeheadline
     \vbox to \pageheight{
       \boxmaxdepth=\maxdepth
       \ifvoid\margin\else 
       \rlap{\kern31pc\vbox to 0pt{\kern4pt\box\margin\vss}}\fi
       \ifvoid\topins\else\unvbox\topins\vskip\skip\topins\fi
       ##1                                  
       \vskip\pagebottomfiller
       \ifvoid\footins\else\vskip\skip\footins\footrule\unvbox\footins\fi
       \ifr@ggedbottom\kern-\dimen@ \vfil\fi}  
       \makefootline}
     \advancepageno\frontpagefalse}
\def\makeheadline{\vbox to\z@{\vskip-22.5\p@
  \hbox to \pagewidth{\vbox to8.5\p@{}\the\headline}\vss}\nointerlineskip}
   \def\makefootline{\baselineskip = 1.5\normalbaselineskip
             \hbox to \pagewidth{\the\footline}}
   \def\footrule{\dimen@=\prevdepth\nointerlineskip
      \vbox to 0pt{\vskip -0.25\baselineskip \hrule width 0.62\pagewidth \vss}
      \prevdepth=\dimen@ }
   \def\Vfootnote##1{\insert\footins\bgroup
      \interlinepenalty=\interfootnotelinepenalty \floatingpenalty=20000
      \singl@true\doubl@false\Tenpoint \hsize=\pagewidth
      \splittopskip=\ht\strutbox \boxmaxdepth=\dp\strutbox
      \leftskip=\footindent \rightskip=\z@skip
      \parindent=0.5\footindent \parfillskip=0pt plus 1fil
      \spaceskip=\z@skip \xspaceskip=\z@skip \footnotespecial
      \Textindent{##1}\footstrut\futurelet\next\fo@t}
 \def\sp@cecheck##1{\dimen@=\pagegoal\advance\dimen@ by -\pagetotal
      \ifdim\dimen@<##1 \ifdim\dimen@>0pt \vfil\break \fi\fi}
 \def\endleftcolumn{\dimen@=\pagegoal\advance\dimen@ by -\pagetotal
      \ifdim\dimen@<\chapterminspace \ifdim\dimen@>0pt \vfil\break \fi
      \hbox{\vbox{\hrule width \columnwidth}\hbox to 0.4pt
      {\vrule height 10pt depth 0pt}\hfil}\fi}
 \def\beginrightcolumn{\dimen@=\pagegoal\advance\dimen@ by -\pagetotal
      \ifdim\dimen@<\chapterminspace \ifdim\dimen@>0pt \vfil\break \fi
      \hbox to \hsize{\hss\hbox{\vrule height 0pt depth 10pt
      \vbox{\hrule width \columnwidth}}}\fi}
}
%
%
%
%
%
%
\newif\ifpr@printstyle \pr@printstylefalse
\newbox\leftpage
\newdimen\fullhsize
\newdimen\titlepagewidth
\newdimen\pagetextwidth
\def\preprintstyle{%
       \message{(This will be printed PREPRINTSTYLE)}
       \let\lr=L
       \frontpagetrue
       \pr@printstyletrue
       \vsize=7truein
       \pagetextwidth=4.75truein
       \fullhsize=10truein
       \titlepagewidth=8truein
       \normalspace
       \Tenpoint
       \voffset=-.31truein
       \hoffset=-.46truein
       \iffrontpage\hsize=\titlepagewidth\else\hsize=\pagetextwidth\fi
 \output={%
    \iffrontpage
      \shipout\vbox{\special{\printertype}\makeheadline
      \hbox to \fullhsize{\hfill\pagebody\hfill}}
      \advancepageno
    \else
       \almostshipout{\leftline{\vbox{\pagebody\makefootline}}}\advancepageno
    \fi}
        \def\almostshipout##1{\if L\lr \count2=1
             \message{[\the\count0.\the\count1.\the\count2]}
        \global\setbox\leftpage=##1 \global\let\lr=R
                             \else \count2=2
        \shipout\vbox{\special{\printertype}
        \hbox to\fullhsize{\hfill\box\leftpage\hskip0.5truein##1\hfill}}
        \global\let\lr=L     \fi}
   \multiply\chapterminspace by 7 \divide\chapterminspace by 9
   \multiply\sectionminspace by 7 \divide\sectionminspace by 9
   \multiply\referenceminspace by 7 \divide\referenceminspace  by 9
   \multiply\chapterskip by 7 \divide\chapterskip  by 9
   \multiply\sectionskip  by 7 \divide\sectionskip  by 9
   \multiply\headskip   by 7 \divide\headskip by 9
   \multiply\baselineskip   by 7 \divide\baselineskip by 9
   \multiply\abovedisplayskip by 7 \divide\abovedisplayskip by 9
   \belowdisplayskip = \abovedisplayskip
\def\advancepageno{\if L\lr \gl@bal\advance\pagen@ by 1\fi
   \ifnum\pagenumber<0 \gl@bal\advance\pagenumber by -1
    \else\gl@bal\advance\pagenumber by 1 \fi
    \gl@bal\frontpagefalse  \swing@
    \gl@bal\hsize=\pagetextwidth}
} 

\tolerance=1000
\def\printertype{ps: }
%

\paperheadline={\ifdr@ftmode\hfil\draftdate\else\hfill\fi}
\def\advancepageno{\gl@bal\advance\pagen@ by 1
   \ifnum\pagenumber<0 \gl@bal\advance\pagenumber by -1
    \else\gl@bal\advance\pagenumber by 1 \fi
    \gl@bal\frontpagefalse  \swing@
    \gl@bal\hsize=\pagetextwidth} 
\def\papersize{\fullhsize=6.5in
               \pagetextwidth=6.5in
               \hsize=\fullhsize
               \vsize=9truein
               \hoffset=0.05 truein
               \voffset=-0.1truein
               \advance\hoffset by\HOFFSET
               \advance\voffset by\VOFFSET
               \pagebottomfiller=0pc
               \skip\footins=\bigskipamount
               \normalspace }
\papers
\def\lettersize{\fullhsize=6.5in
                \pagetextwidth=6.5in
                \hsize=\fullhsize
                \vsize=8.5in
                \hoffset=0in
                \voffset=0.5in
                \advance\hoffset by\HOFFSET
                \advance\voffset by\VOFFSET
                \pagebottomfiller=\letterbottomskip
                \skip\footins=\smallskipamount
                \multiply\skip\footins by 3
                \singlespace }
\def\semi{;\hfil\break}
%
%
\newtoks\chapterheadstyle  \chapterheadstyle={\relax}
\def\chapter#1{{\the\chapterheadstyle\par \penalty-300 \vskip\chapterskip
   \spacecheck\chapterminspace
   \chapterreset \titlestyle{\chapterlabel.~#1}
   \nobreak\vskip\headskip \penalty 30000
   \message{(\the\chapternumber. #1)}
  {\pr@tect\wlog{\string\chapter\space \chapterlabel}} }}

\def\APPENDIX#1#2{{\the\chapterheadstyle\par\penalty-300\vskip\chapterskip
   \spacecheck\chapterminspace \chapterreset \xdef\chapterlabel{#1}
   \titlestyle{APPENDIX #2} \nobreak\vskip\headskip \penalty 30000
   \wlog{\string\Appendix~\chapterlabel} }}
\def\chapterreset{\gl@bal\advance\chapternumber by 1
   \ifnum\equanumber<0 \else\gl@bal\equanumber=0\fi
   \gl@bal\sectionnumber=0 \let\sectionlabel=\rel@x
   {\pr@tect
       \xdef\chapterlabel{{\the\chapterstyle{\the\chapternumber}}}}}%

%
%
\newif\ifdr@ftmode
\newtoks\r@flabeltoks
\def\draftmode{
   \pagetextwidth=6truein
   \fullhsize=6truein
   \titlepagewidth=6truein
   \normalspace
   \hoffset=0.3truein
   \voffset=0.2truein
   \hsize=\pagetextwidth
   \tenpoint
   \baselineskip=24pt plus 2pt minus 2pt
   \dr@ftmodetrue
   \message{ DRAFTMODE }
   \writedraftlabels
   \def\timestring{\begingroup
     \count0 = \time \divide\count0 by 60
     \count2 = \count0  
     \count4 = \time \multiply\count0 by 60
     \advance\count4 by -\count0   
     \ifnum\count4<10 \toks1={0} 
     \else \toks1 = {}
     \fi
     \ifnum\count2<12 \toks0={a.m.} %
          \ifnum\count2<1 \count2=12 \fi
     \else            \toks0={p.m.} %
           \ifnum\count2=12 
           \else
           \advance\count2 by -12 
           \fi
     \fi
     \number\count2:\the\toks1 \number\count4\thinspace \the\toks0
   \endgroup}%
   \def\draftdate{{{\tt preliminary version:}\space{\rm
                                  \timestring\quad\the\date}}}
\def\R@FWRITE##1{\ifreferenceopen \else \gl@bal\referenceopentrue
     \immediate\openout\referencewrite=\jobname.refs
     \toks@={\begingroup \refoutspecials \catcode`\^^M=10 }%
     \immediate\write\referencewrite{\the\toks@}\fi
     \immediate\write\referencewrite%
     {\noexpand\refitem{\the\r@flabeltoks[\the\referencecount]}}%
     \p@rse@ndwrite \referencewrite ##1}
\def\refitem##1{\r@fitem{##1}}
\def\REF##1##2{\reflabel##1 \REFNUM ##1\REFWRITE{\ignorespaces ##2}}
\def\Ref##1##2{\reflabel##1 \Refnum ##1\REFWRITE{ ##2}}
\def\REFS##1##2{\reflabel##1 \REFNUM ##1%
\gl@bal\lastrefsbegincount=\referencecount\REFWRITE{ ##2}}
\def\refs{\REFS\?}
\def\refc{\REF\?}
\let\refscon=\refc       \let\REFSCON=\REF
}
\def\nodraftlabels{\def\leqlabel##1{}\def\eqlabel##1{}\def\reflabel##1{}%
\def\leqlabel##1{}}
\def\writedraftlabels{
  \def\eqlabel##1{{\escapechar-1\rlap{\sevenrm\hskip.05in\string##1}}}%
  \def\leqlabel##1{{\escapechar-1\llap{\sevenrm\string##1\hskip.05in}}}%
  \def\reflabel##1{\r@flabeltoks={{\escapechar-1\sevenrm\string##1\hskip.06in%
}}}}

\nodraftlabels   
\dr@ftmodefalse  
%
%
%

\def\eqn#1{\eqno\eqname{#1}\eqlabel#1}
%
%
\def\eqinsert#1{\noalign{\dimen@=\prevdepth \nointerlineskip
   \setbox0=\hbox to\displaywidth{\hfil #1}
   \vbox to 0pt{\kern 0.5\baselineskip\hbox{$\!\box0\!$}\vss}
   \prevdepth=\dimen@}}  

%


%
%
%
%
%
\def\refout{\par\penalty-400\vskip\chapterskip
   \spacecheck\referenceminspace
   \ifreferenceopen \Closeout\referencewrite \referenceopenfalse \fi
   \line{\ifpr@printstyle\twelverm\else\fourteenrm\fi
         \hfil REFERENCES\hfil}\vskip\headskip
   \input \jobname.refs
   }
\def\ACK{\par\penalty-100\medskip \spacecheck\sectionminspace
   \line{\ifpr@printstyle\twelverm\else\fourteenrm\fi
      \hfil ACKNOWLEDGEMENTS\hfil}\nobreak\vskip\headskip }
\def\tabout{\par\penalty-400
   \vskip\chapterskip\spacecheck\referenceminspace
   \iftableopen \Closeout\tablewrite \tableopenfalse \fi
   \line{\ifpr@printstyle\twelverm\else\fourteenrm\fi\hfil TABLE CAPTIONS\hfil}
   \vskip\headskip
   \input \jobname.tabs
   }
\def\figout{\par\penalty-400
   \vskip\chapterskip\spacecheck\referenceminspace
   \iffigureopen \Closeout\figurewrite \figureopenfalse \fi
   \line{\ifpr@printstyle\twelverm\else\fourteenrm\fi\hfil FIGURE
CAPTIONS\hfil}
   \vskip\headskip
   \input \jobname.figs
   }
\def\masterreset{\begingroup\hsize=\pagetextwidth
   \global\pagenumber=1 \global\chapternumber=0
   \global\equanumber=0 \global\sectionnumber=0
   \global\referencecount=0 \global\figurecount=0 \global\tablecount=0
   \endgroup}
%

%

\def\half{{\textstyle{1\over2}}}

\def\fourth{{\textstyle{1\over4}}}
\def\12{{1\over2}}

\def\sla{\raise.15ex\hbox{$/$}\kern-.57em}
\def\leaderfill{\leaders\hbox to 1em{\hss.\hss}\hfill}
\def\dual{{\,^*\kern-.20em}}
\def\bx{{\vcenter{\hrule height 0.4pt
      \hbox{\vrule width 0.4pt height 10pt \kern 10pt
        \vrule width 0.4pt}
      \hrule height 0.4pt}}}
\def\inner{\,{\vcenter{
      \hbox{ \kern 4pt
        \vrule width 0.5pt height 7pt}
      \hrule height 0.5pt}}\,}
\def\sqr#1#2{{\vcenter{\hrule height.#2pt
      \hbox{\vrule width.#2pt height#1pt \kern#1pt
        \vrule width.#2pt}
      \hrule height.#2pt}}}
\def\rect#1#2#3#4{{\vcenter{\hrule height#3pt
      \hbox{\vrule width#4pt height#1pt \kern#1pt
        \vrule width#4pt}
      \hrule height#3pt}}}

\def\bx{{\vcenter{\hrule height 0.4pt
      \hbox{\vrule width 0.4pt height 10pt \kern 10pt
        \vrule width 0.4pt}
      \hrule height 0.4pt}}}

\def\up#1{\leavevmode \raise.16ex\hbox{#1}}
\def\twiddle{\lower.9ex\rlap{$\kern-.1em\scriptstyle\sim$}}
\def\bigtwiddle{\lower1.ex\rlap{$\sim$}}
\def\gtwid{\mathrel{\raise.3ex\hbox{$>$\kern-.75em\lower1ex\hbox{$\sim$}}}}
\def\ltwid{\mathrel{\raise.3ex\hbox{$<$\kern-.75em\lower1ex\hbox{$\sim$}}}}
\def\square{\kern1pt\vbox{\hrule height 1.2pt\hbox{\vrule width 1.2pt\hskip 3pt
   \vbox{\vskip 6pt}\hskip 3pt\vrule width 0.6pt}\hrule height 0.6pt}\kern1pt}
\def\tdot#1{\mathord{\mathop{#1}\limits^{\kern2pt\ldots}}}

\def\pmb#1{\setbox0=\hbox{#1}    
  \kern-.025em\copy0\kern-\wd0
  \kern  .05em\copy0\kern-\wd0
  \kern-.025em\raise.0433em\box0 }


\def\prl{\journal Phys. Rev. Lett. }

\hyphenation{anom-aly anom-alies coun-ter-term coun-ter-terms}
\def\inv{^{\raise.15ex\hbox{${\scriptscriptstyle -}$}\kern-.05em 1}}

\def\Dsl{\,\raise.15ex\hbox{/}\mkern-13.5mu D} 
\def\dsl{\raise.15ex\hbox{/}\kern-.57em\partial}

\def\boxeqn#1{\vcenter{\vbox{\hrule\hbox{\vrule\kern3pt\vbox{\kern3pt
        \hbox{${\displaystyle #1}$}\kern3pt}\kern3pt\vrule}\hrule}}}
\def\mbox#1#2{\vcenter{\hrule \hbox{\vrule height#2in
                \kern#1in \vrule} \hrule}}  
%

\def\darr#1{\raise1.5ex\hbox{$\leftrightarrow$}\mkern-16.5mu #1}
\def\roughly#1{\raise.3ex\hbox{$#1$\kern-.75em\lower1ex\hbox{$\sim$}}}
%
%
%
\def\ack{\ACK}   
\font\titlerm=cmr10 scaled \magstep 4
\def\TITLEPAGE{\frontpagetrue\pageno=1\pagenumber=1}
%
\def\CALT#1{\hbox to\hsize{\tenpoint \baselineskip=12pt
        \hfil\vtop{\hbox{\strut CALT-68-#1}
        \hbox{\strut DOE RESEARCH AND}
        \hbox{\strut DEVELOPMENT REPORT}}}}



%

\def\WISCONSIN{\vskip15pt\vbox{\hbox{\centerline{\it Department of Physics}}
        \vskip 0pt
  \hbox{\centerline{\it 1150 University Avenue}}\vskip 0pt
  \hbox{\centerline{\it University of Wisconsin, Madison, WI 53706 USA}}}}

\def\TITLE#1{\vskip 1in \centerline{\titlerm #1}}

\def\AUTHOR#1{\vskip .5in \centerline{#1}}

\def\ABSTRACT#1{\vskip .5in \vfil \centerline{\twelvepoint \bf Abstract}
        #1 \vfil}
\def\ENDTITLEPAGE{\vfill\eject\pageno=2\pagenumber=2}

\def\underwig#1{{
\setbox0=\hbox{$#1$}
\setbox1=\hbox{}
\wd1=\wd0
\ht1=\ht0
\dp1=\dp0
\setbox2=\hbox{$\rm\widetilde{\box1}$}
\dimen@=\ht2 \advance \dimen@ by \dp2 \advance \dimen@ by 1.5pt
\ht2=0pt \dp2=0pt
\hbox to 0pt{$#1$\hss} \lower\dimen@\box2
}}
\def\bunderwig#1{{
\setbox0=\hbox{$#1$}
\setbox1=\hbox{}
\wd1=\wd0
\ht1=\ht0
\dp1=\dp0
\setbox2=\hbox{$\seventeenrm\widetilde{\box1}$}
\dimen@=\the\ht2 \advance \dimen@ by \the\dp2 \advance \dimen@ by 1.5pt
\ht2=0pt \dp2=0pt
\hbox to 0pt{$#1$\hss} \lower\dimen@\box2
}}
\def\journal#1&#2(#3){\unskip, \sl #1~\bf #2 \rm (19#3) }
\def\npjournal#1&#2&#3&#4&{\unskip, #1~\rm #2 \rm (#3) #4}
\gdef\prjournal#1&#2&#3&#4&{\unskip, #1~\bf #2, \rm #4 (#3)}

\def\ket#1{\left| #1\right\rangle}

\let\int=\intop         
\catcode`@=12 
\masterreset
\normalspace
\chapterheadstyle={\bf}
\overfullrule=0pt
\def\RPS{reduced phase space}
\def\DGB{Dirac--Gupta--Bleuler}
\def\Reals{{\rm I\!R}}
\def\Oop{{$\widehat{\cal O}$}}
\def\prl{{\it Phys. Rev. Lett.\/ }}
\def\SUBTITLE#1{\vskip 19pt  \centerline{\fourteenrm #1}}
\TITLEPAGE
\rightline{{\tenpoint\baselineskip=12pt
           \vtop{\hbox{\strut MAD/TH-92-05 }
                 \hbox{\strut December 1992 }
                 \hbox{\strut hep-th/9304113} }}}

\TITLE{ Phase Space Reduction and Vortex Statistics }
\SUBTITLE{ An Anyon Quantization Ambiguity }
\AUTHOR{ Theodore J. Allen, Andrew J. Bordner, \it and \rm Dennis B. Crossley }

\WISCONSIN

\ABSTRACT{ We examine the quantization of the motion of two charged vortices in
a Ginzburg--Landau theory for the fractional quantum Hall effect recently
proposed by the first two authors.  The system has two second-class constraints
which can be implemented either in the \RPS\ or \DGB\ formalism.  Using the
intrinsic formulation of statistics,  we show that these two ways of
implementing the constraints are inequivalent unless the vortices are quantized
with conventional statistics; either fermionic or bosonic.}

\ENDTITLEPAGE

\chapter{ Introduction }

The fractional quantum Hall effect is an intriguing example of the effects of
a highly ordered ground state much like superconductivity or superfluidity.
In analogy to the description of these macroscopic quantum states by an
effective field theory, there is a description of the fractional quantum Hall
state by an effective scalar field theory.  The former field theories describe
a Bose condensed state, while the interpretation of the effective field theory
for the fractional quantum Hall state is less straightforward.   One can
account for the fractional conductivity of the state through topological vortex
excitations of the vacuum of the effective field theory which are effectively
fractionally charged.   These vortex excitations correspond exactly to
Laughlin's fractionally charged quasiparticles\Ref\Laughlin{R.B.~Laughlin, \prl
{\bf 50} (1983) 1395.}  in the many-body  theory.  It has been
argued\Ref\Arovas{D.~Arovas, R.~Schrieffer and F.~Wilczek, \prl\  {\bf 53}
(1984) 722.} that Laughlin's quasiparticles are \it anyons\rm, particles having
statistics intermediate between fermions and bosons.  This is a situation
which, for point particles at least, is unique to two spatial dimensions.    An
interesting related question is  the statistics of vortex excitations in
two-dimensional superfluid films,\Ref\VortStat{R.Y. Chiao, A. Hansen and A.A.
Moulthrop, \prl {\bf 54} (1985) 1339\semi A. Hansen, A.A. Moulthrop and R.Y.
Chiao, \prl {\bf 55} (1985) 1431\semi F.D.M. Haldane and Yong-Shi Wu, {\it
Phys. Rev. Lett. } {\bf 55} (1985) 2887.}  which is usually considered in the
\RPS\ formalism.\REF\LM{J.M. Leinaas and J. Myrheim, {\it Nuovo Cimento B}
{\bf 37} (1977) 1.} \REF\Sorkin{R.D. Sorkin, {\it Phys.  Rev. } {\bf D27}
(1983) 1787.} \REF\Bourdeau{M. Bourdeau and R. Sorkin,  \it Phys. Rev. \bf D45
\rm (1992) 687.}

We consider here the quantum statistics of Ginzburg-Landau vortex excitations
within the intrinsic, or topological, formulation.\refmark{\LM-\Bourdeau}
In the intrinsic formulation, the configuration space
of identical particles is smaller than  the classical configuration space
because configurations which are quantum mechanically indistinguishable are
identified.  Quantum statistics in the  intrinsic view generally arise from the
nontrivial topology of the intrinsic configuration space.  Because the
intrinsic configuration space is not a manifold in general, one encounters
problems which do not occur in the usual treatment of quantum
statistics through the permutation of particle labels. The overlap of two or
more particles becomes a boundary of the configuration space making it
necessary to choose the boundary conditions at these points which  preserve the
self-adjointness of the Hamiltonian.    Implementing the intrinsic  formulation
is reasonably straightforward when the particles are described by a
non-singular quadratic Hamiltonian. When constraints are present, however, the
configuration space is not uniquely defined.   In the present case, there are
two different configuration spaces which are relevant.  One is the actual
configuration space before constraints are imposed and the second is the
configuration space resulting from the \RPS\ quantization.  We find that only
in the case of Fermi or Bose statistics do the \RPS\ and \DGB\ methods agree.

\chapter{ The Intrinsic Formulation of Statistics }

The classical configuration space for a system of $N$ identical  particles
moving in Euclidean $m$-space is $\Reals^{Nm}$, identical particles being
classically distinguishable.  Quantum mechanically, however, identical
particles are not distinguishable and the true or intrinsic configuration
space,
$$\eqalign{ Q_{\rm int} &= {\Reals^{Nm}\\ \Delta\over S_N},\cr \Delta &=
\{({\bf x}_1,\ldots,{\bf x}_i,\ldots,{\bf x}_j,\ldots,{\bf x}_N)\,\,  |\,\,
{\bf x}_i = {\bf x}_j\},\cr  &({\bf x}_1,\ldots,{\bf x}_i,\ldots,{\bf x}_N)
\sim ({\bf x}_{\sigma(1)},\ldots,{\bf x}_{\sigma(i)},\ldots,{\bf
x}_{\sigma(N)}), \quad\sigma\in S_N,\cr }\eqn\intconfig$$
is not generally a smooth manifold as it has conical singularities where two or
more particles coincide.\REF\PiOne{L.S. Schulman, {\it Phys. Rev. } {\bf 176}
(1968) 1558 and {\it Techniques and Applications of Path Integration}," (Wiley,
New York, 1981)\semi\^^M J.B. Hartle and J.R. Taylor, {\it Phys. Rev. } {\bf
D1} (1970) 2226\semi\^^M C. Morette-DeWitt, {\it Phys. Rev. } {\bf D3} (1971)
1\semi\^^M J.S. Dowker, {\it J. Phys.} {\bf A5} (1972) 936;\^^M ``Selected
Topics in Quantum Field Theory," Austin lectures (1979) unpublished;\^^M {\it
J. Phys.} {\bf A18} (1985) 2521\semi\^^M   F. Wilczek, \prl {\bf 49} (1982)
957\semi\^^M R.D. Sorkin,  ``Introduction to Topological Geons," in {\it
Topological Properties and Global Structure of Spacetime}, P.G. Bergmann and V.
de Sabbata, eds. (Plenum, New York, 1986) p. 249\semi\^^M Y.-S. Wu, \prl {\bf
52} (1984) 2103;\^^M \prl {\bf 53} (1984) 111\semi\^^M C.J. Isham, in: {\it
Relativity, Groups and Topology II}, B.S. DeWitt and R. Stora, eds. (Elsevier,
Amsterdam, 1984)\semi\^^M R. MacKenzie and F. Wilczek, {\it Int. J. Mod. Phys.}
{\bf A3} (1988) 2827\semi\^^M A.P. Balachandran, ``Wess-Zumino Terms and
Quantum Symmetries, A Review," in {\it Conformal Field Theory, Anomalies and
Superstrings}, B.E. Baaquie, C.K. Chew, C.H. Oh, and K.K. Phua, eds., (World
Scientific, Singapore, 1988);\^^M ``Topological Aspects of Quantum Gravity," in
{\it Particle Physics-Superstring Theory}," R. Ramachandran and H.S. Mani, eds.
(World Scientific, Singapore, 1988);\^^M ``Classical Topology and Quantum
Phases: Quantum Mechanics," in: {\it Geometrical and Algebraic Aspects of
Nonlinear Field Theory}, S. DeFilippo, M. Marinaro, G. Marmo and G. Vilasi,
eds. (Elsevier, Amsterdam, 1989);\^^M ``Classical Topology and Quantum Phases,"
in: {\it Anomalies, Phases, Defects,...}, M. Bregola, G. Marmo, G. Morandi,
eds., (Bibliopolis, Naples, 1990)\semi\^^M A.P. Balachandran, G. Marmo, B.S.
Skagerstam, A. Stern, \sl Classical Topology and Quantum States,\rm\ (World
Scientific, Singapore, 1991)\semi\^^M E.C.G. Sudarshan, T.D. Imbo and T.R.
Govindarajan, {\it Phys. Lett. }  {\bf 213B} (1988) 471\semi\^^M P.O. Horvathy,
G. Morandi, and E.C.G. Sudarshan, {\it Nuovo Cimento} {\bf 11D} (1989)
201\semi\^^M T.D. Imbo, C.S. Imbo and E.C.G. Sudarshan, {\it Phys. Lett. }
{\bf 234B} (1990) 103\semi\^^M T.D. Imbo and J. March-Russell, {\it Phys. Lett.
} {\bf 252B} (1990) 84 and references therein.}\
In general, the configuration space $Q_{\rm int}$ does not have the same
topology as $\Reals^{Nm}$.  The case which interests us here, that of $N$
particles moving in two-dimensional space, has an intrinsic configuration
space, $Q_{\rm int}$, which is not simply connected.  When  $Q_{\rm int}$ is
not simply connected, the wave functions need not be single-valued  and in
general will belong to a unitary irreducible representation of $\pi_1(Q_{\rm
int})$, the first homotopy group.\refmark{\LM,\Sorkin,\PiOne}   This is the
origin of exotic statistics in two dimensions.\REF\Reed{M. Reed and B. Simon,
{\it Methods of Modern Mathematical Physics, Vol I:  Functional Analysis},
(Academic, San Diego, 1980) Ch.\ VIII.}

A technical complication of the intrinsic formulation is that the space
$Q_{\rm int}$ has singularities; boundaries from which probability might leave
the system.  If unitary time evolution is desired, as it is when particle
number is not allowed to change, Stone's theorem\refmark{\Reed} assures us that
the Hamiltonian must be self-adjoint.

For completeness, we review here facts and definitions about self-adjoint
operators, which are explained in greater detail in Ref.\ [\Reed]. Let \Oop\
be
a densely defined operator on a separable Hilbert space $\cal H$.  The operator
\Oop\  is symmetric if and only if $\widehat{\cal O} \subset \widehat{\cal
O}^\dagger$.  That is, the domain of $\widehat{\cal O}^\dagger$ is not smaller
than the domain of \Oop, $D(\widehat{\cal O})\subset  D(\widehat{\cal
O}^\dagger)$, and the operators agree on the domain of \Oop, $\widehat{\cal
O}\psi = \widehat{\cal O}^\dagger\psi$ for all $\psi\in D(\widehat{\cal O})$.

An operator \Oop\  is self-adjoint if and only if $\widehat{\cal O} =
\widehat{\cal O}^\dagger$.  That is, if and only if \Oop\  is symmetric and
$D(\widehat{\cal O}) = D(\widehat{\cal O}^\dagger)$.  It is easy to show that a
self-adjoint operator has only real eigenvalues.  To determine whether a
symmetric operator \Oop\  is self-adjoint, we examine the spectrum of its
adjoint, $\widehat{\cal O}^\dagger$.  If there are no solutions to
$\widehat{\cal O}^\dagger\psi = \pm i\psi$, then \Oop\  is essentially
self-adjoint, that is, its closure $\widehat{\cal O}^{\dagger\dagger}$ is
self-adjoint.  More generally, if \Oop\  is a symmetric operator, let ${\cal
K}_\pm = {\rm Ker}(\widehat{\cal O}^\dagger \mp i)$, $n_\pm = {\rm dim}({\cal
K}_\pm)$.  The kernels ${\cal K}_\pm$ are the deficiency subspaces of \Oop\
and their dimensions, $n_\pm$, are the deficiency indices of \Oop.  If $n_+ =
n_- = n$, then there is an $n^2$-parameter family of self-adjoint extensions of
\Oop.  The domain of the adjoint of \Oop\ is given by
$$ D(\widehat{\cal O}^\dagger) = D(\widehat{\cal O})\oplus {\cal K}_+ \oplus
{\cal K}_-.\eqn\domadj$$
The self-adjoint extensions are characterized by the unitary maps $U:~{\cal
K}_+  \to {\cal K}_-$.  For each such unitary map $U$, one finds a domain on
which \Oop\ is self-adjoint:
$$ D_{U}(\widehat{\cal O}) = \{ \phi + \psi_+ + U\psi_+ |\, \phi \in
D(\widehat{\cal O}), \,\psi_+\in{\cal K}_+
\}.\eqn\sadomain$$
If $n_+ = n_- = 0$, then \Oop\  is essentially self-adjoint.  If $n_+ \neq
n_-$, then \Oop\  is not self-adjoint and possesses no self-adjoint extensions.

\chapter{ Collective Quantum Mechanics of Two Vortices }

Recently,  two of us constructed an effective field theory for the fractional
quantum Hall effect which has charged vortex
excitations.\Ref\AllBord{T.J.~Allen and A.J.~Bordner, {\it Charged Vortex
Dynamics in Ginzburg-Landau Theory of the Fractional Quantum Hall Effect},
UW-Madison preprint MAD/TH-92-02, hep-th/9206073.}  We found that the
collective
coordinate action for $N$ vortex centers ${\bf X}^A$ with integral vorticities
$n_A$, $A=1,\ldots,N$, in the approximation of point vortices is
$$\eqalign{ S_V &=\int dt\,\bigg(\sum_A \pi n_A\rho_0 \epsilon_{ab}X^A_a{\dot
X}^A_b\cr &\phantom{=}\,- \sum_{A < B} 2\pi\alpha n_An_B \ln|{\bf X}^A - {\bf
X}^B|^2\cr &\phantom{=}\,-\sum_A \pi\alpha n_A eB |{\bf X}^A|^2\biggr).\cr}
\eqn\voraction$$
The quantities $\rho_0$ and $\alpha$ are constants arising from the original
effective field theory, $e$ is the electron charge and $B$ is the constant
background magnetic field.  In what follows, we set $\pi\rho_0 =1$,  and
$2\pi\alpha=1$, to make the manipulations more transparent. It is important to
note that the action \voraction\ is first order in time derivatives, and is
thus quite different from an ordinary nonrelativistic point particle action.
There are two very different methods for constructing the quantum mechanics of
such an action.  The first and most straightforward of these is simply to
observe that the two components of each vortex center are canonically
conjugate because the action \voraction\ already has Hamiltonian form: $S =
\int[p\dot{q} - H(p,q)]\,dt$.  This is the reduced phase space quantization,
which can also be arrived at through the Dirac constraint analysis by replacing
the Poisson brackets of the dynamical variables by Dirac brackets. The action
\voraction\ describes {\it one-dimensional} motion for each vortex in the
reduced phase space quantization.  We are free to choose, say, the $X_1$
coordinates of each vortex center to be the configuration variables and then
the $X_2$ coordinates (after an integration by parts) become the momenta
conjugate to the $X_1$'s,
$$ X^A_2 = -{1\over 2n_A}P^A_1.\eqn\RPSmomentum$$
The last term in the Hamiltonian,
$$ H =  \sum_{A < B} 2\pi\alpha n_An_B \ln|{\bf X}^A - {\bf X}^B|^2 + \sum_A
\pi\alpha n_A eB |{\bf X}^A|^2,\eqn\CanHam$$
is a function of the $|{\bf X}^A|^2$ which become one-dimensional harmonic
oscillator Hamiltonians
$$ |{\bf X}^A|^2 = (X^A_1 )^2 + {1\over (2 n_A )^2} (P_1^A)^2.\eqn\RPSHO$$
The states are functions only of the $X_1^A$'s.  In the analysis of two
vortices, the operator representing $|{\bf X}^{(1)}-{\bf X}^{(2)}|^2$  will be
of central importance.  In the \RPS\ quantization it is a harmonic oscillator,
while in the \DGB\ quantization, it is an angular momentum.

The second, more sophisticated, quantization procedure is due to
Dirac.\Ref\Dirac{P.A.M. Dirac, {\it Lectures on Quantum Mechanics},  Yeshiva
University, (Academic Press, New York, 1967).} This is the procedure employed
in Ref.\ [\AllBord].  In the Dirac method, one takes the original configuration
variables ${\bf X}^A$ and introduces their canonical conjugates ${\bf
P}^A=\partial L/\partial{\dot{\bf X}^A}$.  In this case we find directly the
second-class phase-space constraints
$$ \varphi^A_a({\bf P}^A,{\bf X}^A) = P^A_a +  n_A \epsilon_{ab}X^A_b \approx
0.\eqn\constraints $$
The naive canonical Hamiltonian must be modified, by adding to it quantities
which vanish when the constraints do, so that the constraints are preserved
under time evolution.  When second-class constraints are present, the correct
Hamiltonian is often the ``starred'' Hamiltonian,\Ref\HRT{A.J. Hanson, T.
Regge and C. Teitelboim,  {\it Constrained Hamiltonian Systems}, (Accademia
Nazionale dei Lincei, Rome, 1976), p.\ 10.}
$$ \eqalign{H^* &= H - \{ H, \varphi_a\}(\Delta^{-1})^{ab}\varphi_b,\cr
 \Delta_{ab} &\equiv \{ \varphi_a, \varphi_b \},\cr}\eqn\starredH$$
although there is no unique prescription for constructing the correct
Hamiltonian for a given problem.  We will use the notation $H^*$ for any
modified Hamiltonian which preserves the constraints.

In the Gupta-Bleuler quantization prescription the  physical quantum states are
constructed so that all matrix elements of the above constraints vanish in the
physical basis.  The \DGB\ prescription is more restrictive.  In it, we must be
able to divide the second-class constraints into ``creation'' and
``annihilation'' constraints $\overline{\varphi}_a$ and $\varphi_b$
such that the only weakly non-vanishing Poisson brackets are those between a
creation and an annihilation constraint.  Upon quantization, the physical
states are chosen to be those annihilated by the annihilation
constraint operators
$$\hat\varphi_a\ket{\psi_{\rm phys}} = 0.\eqn\DGBcondn$$
In order that physical states evolve into physical states, it is necessary that
the annihilation constraints evolve into themselves,
$$[H,\hat\varphi_a] = \Lambda_{ab}\hat\varphi_b,\eqn\DGBevolve$$
guaranteeing that \DGBcondn\ at time $t=0$ implies the same at all later times,
$$\hat\varphi_a\ket{\psi(t)} = \hat\varphi_a e^{-itH}\ket{\psi(0)} =
e^{-itH}e^{itH}\hat\varphi_a e^{-itH}\ket{\psi(0)} =  e^{-itH}
(e^{it\Lambda})_{ab}\hat\varphi_b\ket{\psi(0)} = 0. \eqn\DGBtime$$

In constructing the states for two identical unit vortices ($n_{(1)} = n_{(2)}
= 1$), it is useful to take center-of-mass and relative coordinates
$$\eqalign{ {\bf X}^{\rm CM} &= \half({\bf X}^{(1)} + {\bf X}^{(2)}),\cr
{\bf X}^{\rm rel} &= {\bf X}^{(1)} - {\bf X}^{(2)},\cr}\eqn\cmrel$$
because the Hamiltonian and the constraints separate;
$$ \eqalign{ H &= H_{\rm CM} + H_{\rm rel} \cr &= eB\,|{\bf X}^{\rm CM}|^2 +
\ln|{\bf X}^{\rm rel}|^2 + \fourth eB|{\bf X}^{\rm rel}|^2,\cr
\varphi^{\rm CM}_a &=
P^{\rm CM}_a + 2 \epsilon_{ab} X^{\rm CM}_b\approx 0,\cr  \varphi^{\rm rel}_a
&= P^{\rm rel}_{a} + \half \epsilon_{ab} X^{\rm rel}_b\approx 0.\cr }\eqn\sep$$
We find that the complex combinations of constraints
$$\eqalign{\varphi^{\rm CM} &= \varphi^{\rm CM}_1 + i\varphi^{\rm CM}_2 \approx
0,\cr \overline\varphi^{\rm CM} &= \varphi^{\rm CM}_1 - i\varphi^{\rm CM}_2
\approx 0,\cr}\quad \eqalign{\varphi^{\rm rel} &=\varphi^{\rm rel}_1 +
i\varphi^{\rm rel}_2\approx 0,\cr \overline\varphi^{\rm rel} &=\varphi^{\rm
rel}_1 - i\varphi^{\rm rel}_2\approx 0,\cr}\eqn\cplxsep$$
can be written simply in terms of $Z = X^{\rm CM}_1 + i X^{\rm CM}_2$,
$\xi = X^{\rm rel}_1 + i X^{\rm rel}_2$ and the combinations of their momenta
$P=\half\left(P_1 - iP_2\right)$,
$\bar{P}=\half\left(P_1 + iP_2\right)$ as follows.
$$\eqalign{ \varphi^{\rm CM} &= -2i(i\bar{P}_{Z} +  Z),\cr
\overline{\varphi}^{\rm CM} &= -2i(iP_{Z} -
\overline{Z}),\cr}\quad\eqalign{\varphi^{\rm rel} &=
-2i(i\bar{P}_\xi + \fourth \xi),\cr \overline{\varphi}^{\rm rel} &=
-2i(iP_{\xi} - \fourth \overline\xi).\cr} \eqn\phiops$$
The Hamiltonian operator must have Poisson brackets with the constraints again
yielding a combination of constraints.   Because the center-of-mass and
relative Hamiltonians are functions of  $|{\bf X}^{\rm CM}|^2$ and $|{\bf
X}^{\rm rel}|^2$ respectively, we may write the correct Hamiltonians as
functions of the starred variables $|{\bf X}^{\rm CM}|^{2*}$ and  $|{\bf
X}^{\rm rel}|^{2*}$, which have Poisson brackets with each constraint in the
basis \phiops\ proportional to that constraint.
$$\eqalign{|{\bf X}^{\rm CM}|^{2*} &= |{\bf X}^{\rm CM}|^{2} + \fourth \{|{\bf
X}^{\rm CM}|^{2},\varphi^{\rm CM}_a\}\,\epsilon_{ab}\,\varphi^{\rm CM}_b,\cr
|{\bf X}^{\rm rel}|^{2*} &= |{\bf X}^{\rm rel}|^{2} + \{|{\bf X}^{\rm
rel}|^{2},\varphi^{\rm rel}_a\}\,\epsilon_{ab}\,\varphi^{\rm rel}_b. \cr}
\eqn\starvar$$
The Hamiltonian as an explicit function of the starred variables is
$$ \eqalign{ H^*_{\rm CM} &= H_{\rm CM}(|{\bf X}^{\rm CM}|^{2*}) =  { eB\over
2}\bigl(iZP_Z  - i\overline{Z} \bar{P}_{Z}\bigr),\cr
H^*_{\rm rel} &=  H_{\rm rel}(|{\bf X}^{\rm rel}|^{2*}) =
\ln[2(i\xi P_\xi - i\overline{\xi}\bar{P}_\xi)] + { eB \over
2 } \bigl(i\xi P_\xi - i\overline{\xi}\bar{P}_\xi\bigr).\cr}
\eqn\Hops$$
To go to quantum operators, we simply replace the momenta by their derivative
forms
$$\eqalign{ \widehat{\bar{P}}_Z &= -i\partial_{\overline{Z}},\cr
            \widehat{P}_Z &= -i\partial_Z,\cr}\quad
\eqalign{ \widehat{\bar{P}}_{\xi} &= -i\partial_{\overline{\xi}},\cr
          \widehat{P}_\xi &= -i\partial_\xi.\cr}\eqn\Pops$$

The physical states are those which are annihilated by the constraints
$\hat{\varphi}^{\rm rel}= -2i(\partial_{\overline{\xi}} + \fourth\xi)$  and
$\hat{\varphi}^{\rm CM}= -2i(\partial_{\overline{Z}} + Z)$. These states have
the form
$$\Psi(\xi,Z)=F(\xi)e^{-{1\over 4}|\xi|^2}\,G(Z)e^{-|Z|^2},\eqn\twovorstate$$
where $F$ and $G$ are holomorphic functions which do not grow too fast at
infinity so that the states are normalizable.

\chapter{ Vortex Statistics }

The fact that vortices are indistinguishable quantum objects must be built into
the quantum mechanics of several vortices.   When two identical vortices are
present there is no restriction on the center-of-mass coordinate but, from the
intrinsic viewpoint, we must identify relative configurations differing
only by the exchange of the vortices.

In the reduced phase space quantization the relative coordinate $X^{\rm rel}_1$
is restricted to non-negative values. It is important to note that the
condition $X^{\rm rel}_1 = 0$ does not imply that the vortices are actually
coincident.  Coincidence also requires that $X^{\rm rel}_2 = 0$, so it is
reasonable to expect that in the reduced  phase space formalism there should be
no probability loss at $X^{\rm rel}_1 = 0$.   Self-adjointness of the
Hamiltonian, which guarantees this, will require that a boundary condition be
put on the wave functions at $X^{\rm rel}_1 = 0$.

The relative configuration space of two vortices in the \RPS\
quantization is the half-line,
$$ Q_{\rm RPS} = \Reals^+, \eqn\RPSconfig$$
and the Hamiltonian,
$$ H = \ln\bigl(x^2 + p^2\bigr) + {eB\over 4}\bigl(x^2 + p^2\bigr),\eqn\RPSH $$
is a function of the positive operator
$$\widehat{\cal O} = x^2 + p^2.\eqn\rpsop$$
Here we use $x$ as the relative coordinate ${X}^{\rm rel}_1$ and $p$ for
its canonical momentum, $-{X}^{\rm rel}_2$.\REF\ReedSimon{M. Reed and B. Simon,
{\it Methods of Modern Mathematical Physics, Vol II: Fourier Analysis,
Self-Adjointness}, (Academic, San Diego, 1975).}\
At the configuration space boundary, $x=0$, we can use the von Neumann
theory\refmark{\Reed,\ReedSimon} outlined in section 2 to find the conditions
required to guarantee that no probability leaks away.

Because it is a positive operator, it is sufficient to require that
$\widehat{\cal O}$ be self-adjoint.  We start from the domain
$$ {\cal D}^{(0)}(\widehat{\cal O}) = \{ \psi\in L^2(\Reals^+)\, |\, \psi(0)  =
\psi'(0) = 0\}. \eqn\Dnaught$$
Let us denote the eigenvalue of $\widehat{\cal O}$  as $2\lambda +1$.  The
normalizable eigenfunctions of $\widehat{\cal O}^\dagger$ on the half line are
$\psi_\lambda(x) = e^{-x^2/2}H_\lambda(x)$, where
$$\eqalign{ H_\lambda(x) =
{2^\lambda\over\sqrt{\pi}}\biggl[&\cos({\pi\lambda\over2})\Gamma({\textstyle
{\lambda\over 2}} + \half){}_1\!F_1(-{\textstyle{\lambda\over 2}},\half;x^2)\cr
&+ 2x\sin({\pi\lambda\over2})\Gamma({\textstyle{\lambda\over 2}} + 1)
{}_1\!F_1(\half - {\textstyle{\lambda\over 2}}, {\textstyle {3 \over
2}};x^2)\biggr]\cr}\eqn\Hermite$$
is a combination of confluent hypergeometric functions ${}_1\!F_1$ which
reduces to the ordinary Hermite polynomials when $\lambda$ is a non-negative
integer.\Ref\Lebedev{N.N. Lebedev, {\it Special Functions and Their
Applications,} (Dover, New York, 1972) pp.~283-293.} For the eigenfunctions
$\psi_\lambda(x)$ to be normalizable on the whole line, it is necessary that
$\lambda$ be a non-negative integer.  There is no such restriction if
normalizability on the positive half line is the only requirement. We can see
directly from \Hermite\ that the deficiency indices of $\widehat{\cal O}$ on
the domain \Dnaught\  are $n_+=n_-=1$ and thus, that there is a one-parameter
family of self-adjoint extensions to $\widehat{\cal O}$.  These self-adjoint
domains are parametrized by a single real number, $\Theta$.
$${\cal D}_\Theta(\widehat{\cal O}) = \{ \psi\in L^2(\Reals^+) \,\,|\, \,\,\,
\psi'(0) = \tan(\half\Theta)\,\psi(0)\, \}, \quad -\pi<\Theta\leq\pi.
\eqn\SAdomains$$
For each value of $\Theta$ we can solve for the spectrum of $\widehat{\cal O}$.

Imposing the self-adjointness condition \SAdomains\ on the eigenfunctions
$\psi_\lambda(x) = e^{-x^2/2}H_\lambda(x)$, we find a relation  determining the
eigenvalues, $2\lambda +1$, of $\widehat{\cal O}$,
$$ 2\tan\left({\pi\lambda\over2}\right)\,{\Gamma({\lambda\over2} + 1)\over
\Gamma({\lambda\over2} + {1\over 2})} =  \tan(\half\Theta).\eqn\SAspec$$
For $\Theta = 0$ or $\pi$, the spectrum of $\widehat{\cal O}$  can be
determined immediately from eq.\ \SAspec. The spectrum is $2\lambda +1$ with
$\lambda = 2n$, $2n+1$ respectively, $n=0,1,2,\ldots$ . Other values of the
parameter $\Theta$ yield spectra which are not evenly spaced, although they
become so asymptotically in $\lambda$,
$$ \lambda\simeq 2n + {\tan({\Theta\over2})\over\pi\sqrt{n}},\quad n >\!\!>
{\tan^2({\Theta\over2})\over \pi^2},\quad \Theta\neq\pi.\eqn\asymptot$$

We note that only in the two special situations, $\Theta = 0$ or $\pi$, when
the spectrum is a subset of the harmonic oscillator spectrum on the whole line,
can the wave functions be extended smoothly to normalizable functions on the
whole line. In the case $\Theta=0$ the eigenstates can be extended to even
functions while in the case $\Theta = \pi$ they can be extended to  odd
functions on the line.  These are the cases of bosonic or fermionic vortices,
respectively.  This is clear because the wave functions are symmetric or
antisymmetric under the exchange $x\to -x$.  There is no such easy
characterization of the states for general $\Theta$, although, for a related
problem, Hansson, Leinaas and Myrheim\REF\Hansson{T.H. Hansson, J.M. Leinaas,
and J. Myrheim, {\it Nucl. Phys. } {\bf B 384} (1992) 559.}\ have argued that
such states be interpreted as anyonic states.\refmark{\LM,\Hansson}   If we try
to extend the general $\Theta$ states smoothly to the whole line,  we find that
they are are not normalizable.

In the \DGB\ quantization, the relative configuration space is the cone
$$ Q_{\rm DGB} = {\Reals^2\\ \{0\}\over \sim},\quad
(X^{\rm rel}_1,X^{\rm rel}_2)\sim(-X^{\rm rel}_1,-X^{\rm rel}_2),
\eqn\DGBconfig$$
which we can parametrize by polar coordinates
$$Q_{\rm DGB} = \{ (r,\phi)\,|\,0<r<\infty,\> 0\leq\phi<\pi,\>\xi=re^{i\phi}
 \}. \eqn\DGBparam$$
Because the configuration space $Q_{\rm DGB}$ is not simply connected, it  is
not necessary that the state function be single-valued on it, but only that its
modulus $|\Psi(r,\phi)|^2 = |\Psi(r,\phi+\pi)|^2$ be single-valued.
Technically speaking, we require that the state function be a section of a
$U(1)$ bundle over the configuration space $Q_{\rm DGB}$.  That is,
$$\Psi(r,\phi + \pi) = e^{i\theta}\Psi(r,\phi), \quad 0 \leq \theta < 2\pi,
\eqn\Bundle$$
or, equivalently, that as a function of complex variables $\xi$ and
$\overline{\xi}$, it have the monodromy
$$ \Psi(e^{i\pi}\xi,e^{-i\pi}\overline{\xi}) =
e^{i\theta}\Psi(\xi,\overline{\xi}). \eqn\xiBundle$$

According to eq.\ \Hops, the Hamiltonian is again the same  function of an
operator $\widehat{\cal O}$,
$$\widehat{\cal O}=\widehat{\cal O}_{\rm DGB} = 2(\xi\partial_\xi -
\overline{\xi}\partial_{\overline{\xi}})  =
-2i{\partial\over\partial\phi},\eqn\DGBop$$
whose self-adjoint extensions we wish to find.  We note that, compared to its
classical expression, the operator \DGBop\ has an ordering ambiguity
and that in order for there to be any possibility of
agreement with the \RPS\ method we must add the normal-ordering constant $1$ to
$\widehat{\cal O}_{\rm DGB}$.   The  Hilbert space of states now is quite a bit
smaller than all normalizable states. Instead of reducing the configuration
space, the constraints now determine the physical Hilbert space
$${\cal H}_{\rm phys} = {\rm Ker}(\hat{\varphi}^{\rm rel}) = \{ \Psi \in
L^2(Q_{\rm DGB}), \, \Psi(r,\phi + \pi) = e^{i\theta} \Psi(r,\phi) \,\,|\,\,
\Psi({\bf 0}) < \infty,\, \hat{\varphi}^{\rm rel}\Psi = 0\}.\eqn\DGBstates$$
The inner product is the usual one.  We have introduced the condition that the
wave function be regular at the origin, though it is not required for
normalizability, in order that the operator \Oop\ be positive and the
Hamiltonian have real eigenvalues. The condition in eq.\ \DGBstates\ is a
very strong analyticity requirement.   Because of this requirement, the
deficiency indices are $n_+ = n_- = 0$. That is, $\widehat{\cal O}_{\rm DGB}$
is essentially self-adjoint on ${\cal H}_{\rm phys}$. The existence of the
anyonic states follows directly from the condition that an exchange of the
vortices leave the state vector unchanged up to a phase.   Using the normal
ordered operator $\widehat{\cal O}_{\rm DGB} = -2i{\partial\over\partial\phi} +
1$,  and eq.\ \xiBundle, we find
$$e^{i\pi\widehat{J}}\Psi = e^{i\pi(\widehat{\cal O}_{\rm DGB}-1)/2 }\Psi =
\Psi(e^{i\pi}\xi,e^{-i\pi}\overline{\xi}) =
e^{i\theta}\Psi(\xi,\overline{\xi}), \eqn\statistics$$
implying directly the spectrum of $\widehat{\cal O}_{\rm DGB}$,
$$\widehat{\cal O}_{\rm DGB}\ket{n}_\theta = \bigg(2(2n +
{\theta\over\pi})+1\bigg)\ket{n}_\theta,\quad n=0,1,2,3,\ldots\>.
\eqn\DGBspectrum$$
When $\theta= 0$, the states are bosons, and when $\theta = \pi$ the states are
fermions.  For all values of $\theta$ the eigenvalues are evenly spaced, while
the eigenvalues in the \RPS\ quantization are only evenly spaced for bosons or
fermions.  Up to a normal-ordering constant, the spectrum of $\widehat{\cal O}$
is the same in both quantizations as long as the vortices are taken to be
either fermions or bosons.

\chapter{ Conclusions }

In the intrinsic formulation of quantum statistics,  we have found that the
\RPS\ and \DGB\ quantizations of identical vortices are equivalent (in the
sense that the observables have identical spectra) only when the vortices are
quantized with conventional Bose or Fermi statistics. In each case the relative
Hamiltonian is given by
$$ H = \ln(\widehat{\cal O}) + {eB\over 4}\widehat{\cal O},\eqn\GenHam $$
but the specific operators $\widehat{\cal O}$ are different.  In the \RPS\
quantization, $\widehat{\cal O}$ has a one parameter set of self-adjoint
extensions which determine the vortex statistics, while in the \DGB\
quantization $\widehat{\cal O}$ is essentially self-adjoint and the quantum
statistics arise from the topology of the configuration space.  The statistics
parameter comes in through the choice of a specific $U(1)$ bundle.

Following Refs.\ [\LM] and [\Hansson], we might identify vortices in the \RPS\
quantization with the most general boundary conditions as anyons, although this
is a delicate issue as we observe that the states in the \RPS\ quantization can
be chosen real and are therefore invariant under time reversal, while the
states (and the Hamiltonian) in the \DGB\ quantization are not time-reversal
invariant because they cannot be real. The time reversal invariance in the
\RPS\ quantization results from the fact that the Hamiltonian is a real
function of the squares of the dynamical variables and not of the variables
themselves.  Thus there is a loss of some phase information.  In particular, it
is impossible to know whether the relative coordinate $X^{\rm rel}_1$ leads or
lags $X^{\rm rel}_2$ since this information is lost when the circular motion of
the vortices is projected onto a single line.  Besides preserving phase
information, the \DGB\ quantization is also preferable if we wish to interpret
anyonic vortices as Laughlin quasiparticles in the  fractional quantum Hall
effect because there is an exact correspondence between states \twovorstate\
and the Laughlin state.

\ack

We thank A.P.~Balachandran for suggesting this investigation,  L.~Durand for
the solution to the harmonic oscillator on the half-line, and J.~Robbin and
S.~Angenant for useful discussions. This work was supported in part by DOE
grant No.\ DE-AC02-76-ER00881.

\refout
\bye